\documentclass{article}

\usepackage{arxiv}

\usepackage[utf8]{inputenc} 
\usepackage[T1]{fontenc}    
\usepackage{hyperref}       
\usepackage{url}            
\usepackage{booktabs}       
\usepackage{amsmath,amssymb,amsfonts}       
\usepackage{nicefrac}       
\usepackage{microtype}      
\usepackage{lipsum}		    
\usepackage{graphicx}
\usepackage{natbib}
\usepackage{doi}
\usepackage{multirow}%
\usepackage{amsthm}%
\usepackage{mathrsfs}%
\usepackage[title]{appendix}%
\usepackage{xcolor}%
\usepackage{textcomp}%
\usepackage{manyfoot}%
\usepackage{algorithm}%
\usepackage{algorithmicx}%
\usepackage{algpseudocode}%
\usepackage{listings}%

\title{Testing Preferential Sampling}

\author{ \href{https://orcid.org/0000-0001-6020-9373}{\includegraphics[scale=0.06]{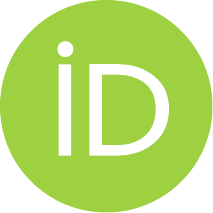}\hspace{1mm}Isabel~Nat\'{a}rio}\thanks{Corresponding author.} \\
	Department of Mathematics of the Nova School of Science and Technology\\
	Center for Mathematics and Applications (NOVA Math)\\
	NOVA University of Lisbon, Caparica, Portugal \\
	\texttt{icn@fct.unl.pt} \\
	\And
	\href{https://orcid.org/0000-0001-9025-2962}{\includegraphics[scale=0.06]{orcid.pdf}\hspace{1mm}Andreia~Monteiro} \\
	Instituto Polit\'{e}cnico do C\'{a}vado e Ave (IPCA)\\
	Barcelos, Portugal \\
	\texttt{andreiaforte50@gmail.com} \\
}

\renewcommand{\shorttitle}{Testing Preferential Sampling}

\hypersetup{
pdftitle={A template for the arxiv style},
pdfsubject={q-bio.NC, q-bio.QM},
pdfauthor={David S.~Hippocampus, Elias D.~Striatum},
pdfkeywords={First keyword, Second keyword, More},
}

\begin{document}
\maketitle

\begin{abstract}
Geostatistics aims to infer a spatially continuous phenomenon from observations collected at a finite number of locations, frequently measured with error.  Whenever there is stochastic dependence between the spatial and sampling processes, preferential sampling occurs. Ignoring this problem drives to incorrect and biased estimates and, therefore, recognizing it is quite important, but not always simple to execute and understand. In this work, a test for assessing preferential sampling, simple and easy to implement, is presented, overcoming the previous concerns. It is based on the dependence between the number of sampled points and the values of the corresponding measures. The  performance of the proposed test id assessed through a large simulation study, which consideres different levels of preferentiability, relation with a covariate, different sample sizes and different test procedure conditions. The results are quite encouraging, with high levels of correct preferential sampling detections, further confirmed by the test application to already known real data sets of lead concentrations in moss samples and red and blue shrimp capture data.

\end{abstract}


\section{Introduction}\label{sec1}

Geostatistics models phenomena with an inherent spatial nature, each described by a continuous stochastic process in a certain spatial domain $\cal{D}$, based on observations sampled in a predetermined set of points (fixed or randomly chosen) of $\cal{D}$, aiming to estimate the spatial phenomenon in the whole domain. A common problem that has been identified in geostatistics is the use of preferential sampling data, whenever there is an unwanted association between the spatial stochastic process to be modelled and the sampling process of the data collected for that purpose \citet{DiggleEtal2010}. This is frequent when data are collected for  reasons other than of estimation, not following a proper statistical sampling design, such as when using fishery data for estimating species abundance \citet{SimoesEtal2023}, or air pollution data collected in monitoring stations that are located  where the pollution levels are expected to be higher, for estimating air pollution distribution \citet{ShaddickZidek2014}.

Being conditioned on either fixed or randomly selected locations, the model based approaches that are commonly used for geostatistical data, \citet{DiggleRibeiro07}, when used with preferential sampled data, can result in quite biased predictions of the spatial process, \citet{Watson2021}, as this type of data may correspond to a reduced variability of the response variable and also the chosen locations may carry relevant information about the response variable, \citet{Dinsdale2019}.
This problematic issue for geostatistical data can be taken into account in the modelling itself, mitigating the biases it causes, \citet{DiggleEtal2010, ShaddickZidek2014, Dinsdale2019}, or it might also be addressed by the inclusion of covariates that, by also being associated with the underlying spatial stochastic process, may be able to partially remove the effects of the PS on the estimation, \citet{GelfandEtal12, Illian2012toolbox, Illian2013, Raeisi2021}. In both cases, the stochastic dependence between the locations of data and the stochastic process to be modelled, induced by preferential sampling, is taken into account.

Thus, it is of the utmost importance to be able to identify if geostatistical data is preferentially sampled, preferably in a routine way.
This has been recognized by a number of authors, although not entirely in satisfactory or global manner. Common to all proposals is a point process approach. Schlather et al., \citet{Schlather2004}, developed a Monte-Carlo independence test between the location of the points, viewed as a marked point process, and the Gaussian marks, representing the spatial phenomena under study modelled as a Gaussian random process. The test is based on two characteristics of marked point processes, established by the authors, namely the conditional expectation and variance of a mark, given the existence of another point of the process at a distance $t$. Guan and Afshartous, \citet{GuanAfshartous2007}, following Guan, \citet{Guan2006}, develop an analytic test (not requiring simulation) based on the same conditional expectation and variance of a mark, that does not assume the normality of the marks. For this, the considered test statistics is based on a kernel-type of estimator of the aforementioned conditional expectation that is compared for different $t$ values, and that ought to be similar under the independence hypothesis. In order to keep the test non-parametric, a subsampling method is employed to estimate the covariance of the test statistics that requires the study region to be divided into a reasonable number of similar non-overlapping areas, assumed to be independent. This results into an approximate qui-square test, that requires quite large samples sizes as well as the subregions themselves, which may constitute a difficulty when the spatial range is large in comparison to the study domain. Both tests assume stationarity and isotropy. Another possibility for identifying preferential sampling when addressing it through modelling, is considering a joint model for the marks and the points conditional on their possible mutual dependence on the common spatial process deriving the marks, weighted by a fixed parameter indicator of the preferentiability if different from zero, \citet{DiggleEtal2010, Dinsdale2019, Pennino2019}. Recognizing the complexity of these approaches, Watson proposed another test for preferential sampling, using nearest neighbour distances to assess local clustering of points. A significant correlation between
the nearest neighbour distances and the marks at each of sampled locations indicates preferential sampling. A point process model is fitted to the data, under the null hypothesis of no preferential sampling, and Monte Carlo simulations are used to compute empirical p-values. A strong correlation in the observed data compared to simulations suggests evidence of preferential sampling. This test does not have distributional requirements on the response (marks) and is available in a R package.

All the above mentioned tests have some drawbacks to their use. The Schlather et al. approach, \citet{Schlather2004}, requires Gaussian marks, not generalizing to non-continuous marks; Although this is not required by the Guan approach \citet{Guan2006, GuanAfshartous2007}, as their approach does not require to fit a parametric model to the marks, it needs large samples sizes as well as suitable subregions of the study domain, which is often problematic. As for the Watson test, \citet{Watson2021}, not only it requires modeling the intensity function of the spatial point process, which involves choosing a theoretical model such as Poisson, Cox, etc., which might constitute a challenge for more applied practitioners, as well as there are some computational limitations. Although the test is described as fast, its application on large data sets or in contexts with high dimensionality can be computationally intensive, especially when multiple Monte Carlo simulations are required.

To address these limitations, we propose a simple, non-parametric test that can be applied to any geostatistical dataset without requiring model fitting or distributional assumptions. The idea behind the test is that partitioning the area of interest in regular disjoint subregions and considering in each of those the number of observed points and the average value of the marks of those points, in case of preferential sampling these measures should be associated. This association is assessed using Spearman’s rank correlation, either in its classical frequentist form or through a Bayesian formulation based on latent normal scores. The Bayesian version is particularly appealing in spatial applications, as it provides a full posterior distribution for the correlation and a Bayes factor for hypothesis testing.

This paper organizes as follows. In Section \ref{NETAL:sec2} the proposed test is described, with two variations. In Section \ref{NETAL:sec3} a simulation study is run to illustrate the test and in Section \ref{NETAL:sec4} the test is applied to real data on lead concentrations in moss samples \cite{DiggleEtal2010,Watson2021} and red and blue shrimp fishery data  \cite{Pennino2019}, previously analysed for preferential sampling. Finally, Section \ref{NETAL:sec4} discusses results.

%

\section{The MLC test}\label{NETAL:sec2}

This work presents a test for preferential sampling on Geostatistics, for assessing wether the sampling locations $\pmb{X}=(X_1,\ldots,X_n)$ selected to monitor a spatial continuous stochastic process $\{S(x)$, $x \in \cal{D}\}$, depend stochastically on the process they are measuring.
Typically, $\{S(x)$, $x \in \cal{D}\}$ is not observed directly and, instead, noisy observations $Y$ are taken, $\pmb{Y}=(Y_1,\ldots,Y_n)$, $Y_i=S(X_i)+\varepsilon_i$, $i=1,\ldots,n$. The noise is due to measurement error (e.g., nugget effect), \citet{DiggleEtal2010}, captured by $\varepsilon$.

Recalling that we are interested in testing the randomness of the sampling process, which does not hold in case of dependence between the sampling locations and the spatial process, this drives the rational for the test proposed here: if there is dependence between the process of the sampling locations $\pmb{X}$ and the spatial process $\{S(x)$, $x \in \cal{D}\}$ - and thus its measurements $\pmb{Y}$ -  than it is expected that, considering a partition of the area of interest, the number of sampled points in each area of the partition is correlated with the values of the corresponding measurements. Thinking of this as $\pmb{X}$ being a marked point process with marks $\pmb{Y}$, it is expected that the points $\pmb{X}$ and the marks $\pmb{Y}$ are dependent in the case of preferential sampling.

Following this line of reasoning, considering the domain of interest $\cal{D}$, the sampling locations $\pmb{X}=(X_1,\ldots,X_n)$ and the observations $\pmb{Y}=(Y_1,\ldots,Y_n)$, the test begins by considering a regular partition of $\cal{D}$ formed by $d$ squares of equal size and computes the number $N_{P,j}$ of sampling points within the $j$th square, $j=1,\ldots,d$, and the mean value of the observations within the $j$th square, $\bar{Y}_j$, $j=1,\ldots,d$. The test can then be taken, performing a Spearman correlation test between the number of points in each cell $\{ N_{P,j}, j=1,\ldots,d \}$ and the average measures in each cell $\{\bar{Y}_j, j=1,\ldots,d\}$, corrected for ties. This test can be considered both under the frequentist, \citet{Spearman1904}, or bayesian framework, \citet{vanDoorn2020}, being the latter more commonly used for spatial modelling. We coined this the Means and Locations Correlation test, MLC test.

The square side size $l$ can be data and/or domain dependent and is preferably considered to be the square root of the average area per point, $l=\sqrt{\mbox{Area}_{\cal{D}}\over n}$, where Area$_{\cal{D}}$ represents the total area of the domain of interest and $n$ is the number of sample points. Depending on this number $n$, which in environmental applications tends to be high, this choice ensures a sufficiently large number of data points to run the Spearman test in its approximate frequentist version and, more accurately, in its bayesian version \citet{Natario2024}.
Other choices of the grid side size are possible, such as $l={h \over 12}$, where $h$ is the maximum distance between all the $n$ sampled points, and these choices are considered and compared in the simulation study presented in section \ref{NETAL:sec3}.

We shall refer indistinctively to the null hypothesis of non-preferential sampling (against the alternative hypothesis of preferential sampling) and the null hypothesis of independence between sampling locations $\pmb{X}$ and measurements of the stochastic process under consideration $\pmb{Y}$ (against the alternative hypothesis of dependence).

\subsection{Frequentist Spearman Independence Test}

Let $(X_1,Y_1),\ldots,(X_n,Y_n)$ be paired i.i.d. data on $X$ and $Y$ measured on the same items and let $r^{x}=(r_1^{x},\ldots,r_n^{x})$ and $r^{y}=(r_1^{y},\ldots,r_n^{y})$ denote their corresponding individual ranks values. The Spearman's correlation $\rho_S$ is given by the Pearson correlation of the ranks $r^{x}$ and $r^{y}$ and is estimated by:

\begin{eqnarray}
r_S=r(r^{x},r^{y})&=&{\sum_{i=1}^n(r_i^{x}-\bar{r}_x)(r_i^{y}-\bar{r}_y) \over \sqrt{\sum_{i=1}^n(r_i^{x}-\bar{r}_x)^2}\sqrt{\sum_{i=1}^n(r_i^{y}-\bar{r}_y)^2}}= \nonumber \\
&& \nonumber \\
&=&1-{6\sum_{i=1}^n(r_i^{x}-r_i^{y})^2 \over n(n^2-1)}
\end{eqnarray}

Testing independence between the random variables that provided the data corresponds to test whether Spearman's correlation $\rho_S$ is null,
\begin{equation}
H_0: \rho_S =0 ~~ vs ~~ H_1: \rho_S \ne 0.
\label{NatarioEtal:NullHip}
\end{equation}

For really small sample sizes ($n<10$, in \texttt{R}), a permutation test using $r_S$ as test statistics, calculates the test p-value as the probability, under the null hypothesis, of obtaining values larger than the absolute value of the observed $r_S$. For larger samples, but yet not so large ($n<1290$, in \texttt{R}), the test p-value can be obtained through an algorithm (AS 89 algorithm, \citet{BestRoberts1975}) that uses an Edgeworth series approximation of (a function of) the test statistics. For really large sample sizes, the following asymptotic $t$ transformation of the test statistics can be used, \citet{Kendall1948Vol1, Kendall1961Vol2}:

$$
t=r_S \cdot \sqrt{n-2 \over 1- r_S^2} ~~ \begin{array}{c}
                                           \mbox{\tiny H$_0$} \\
                                           \sim \\
                                           {\color{white}.}
                                         \end{array}  ~~ t_{(n-2)}
$$

\subsection{Bayesian Spearman Independence Test}

Spearman's rank test for (\ref{NatarioEtal:NullHip}) is considered in the Bayesian paradigm by van Doorn et al. (2020), \citet{vanDoorn2020}, basing their approach on latent normal models, transforming an ordinal into a parametric problem. Unobserved latent normal scores $z^x$ and $z^y$, associated with the ranks $r^x$ and $r^y$, respectively, are considered, dependent on parameter $\rho_S$. The observed ranks are related to these scores in such a way that the ordinal information is preserved, through the data model,
$$
f(r^x,r^y|z^x,z^y).$$
Representing $\pi(z^x, z^y | \rho_S)$ the latent prior structure, the prior model is given by:
$$
\pi(z^x, z^y | \rho_S) \times \pi(\rho_S).
$$
For $\pi(z^x, z^y | \rho_S)$ a centered bivariate normal density with unit variances and correlation $\rho_S$ is considered. The prior distribution for $\rho_S$ is chosen to be an uniform distribution.

The authors have developed an MCMC sampler, \citet{vanDoorn2020}, allowing to obtain the posterior distributions of the scores and the Spearman's correlation, from which the independence test (\ref{NatarioEtal:NullHip}) can be performed, based on the Bayes factor for decision making, \citet{Natario2024}. Choosing the Bayes factor $B_{10}$ in favour of hypothesis $H_1$, the preferential hypothesis, \cite{Kass1995} suggest a classification of very strong evidence in favour of preferential sampling for $B_{10}>150$, strong evidence if $20 < B_{10} < 150$, positive evidence if $3 < B_{10} < 20$ and not worth a bare mention evidence in favour of preferential sampling, but still supporting it, if $1 < B_{10} < 3$.

\section{Simulation Study}
\label{NETAL:sec3}

A simulation study was conducted for investigating the properties and performance of the MLC test for preferential sampling, under controlled conditions. For that purpose, it was necessary to simulate realizations of a spatial process described by a Mat\'{e}rn field, i.e., a Gaussian process ruled by a Mat\'{e}rn covariance function. For two locations separated by a distance $u$, this covariance is given by \citep{BlangiardoCameletti2015}:

\begin{align*}
&C(u)={\sigma^2 \over \Gamma(\lambda)2^{\lambda-1}} \left (\kappa\, u \right )^{\lambda} K_{\lambda} \left (\kappa\,u \right ),&
\end{align*}
where $\sigma^2$ is the marginal variance, $K_{\lambda}$ is the modified Bessel function of the second kind and order $\lambda$, $\lambda$ is associated with the field smoothness and is usually fixed to $1$, $\kappa$ is a scale parameter that controls the correlation decay with distance and relates empirically with the range $r$\footnote{The distance at which spatial correlation is almost null.} as $r={\sqrt{8\lambda} \over \kappa}$.

\begin{figure}
\centering
\includegraphics[width=0.475\linewidth, trim=1cm 1.2cm 0cm 0.5cm, clip]{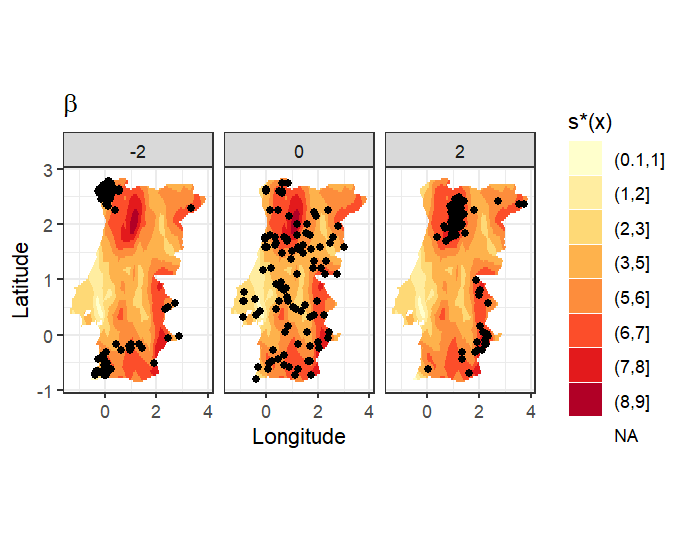}  
\includegraphics[width=0.475\linewidth, trim=1cm 1.2cm 0cm 0.5cm, clip]{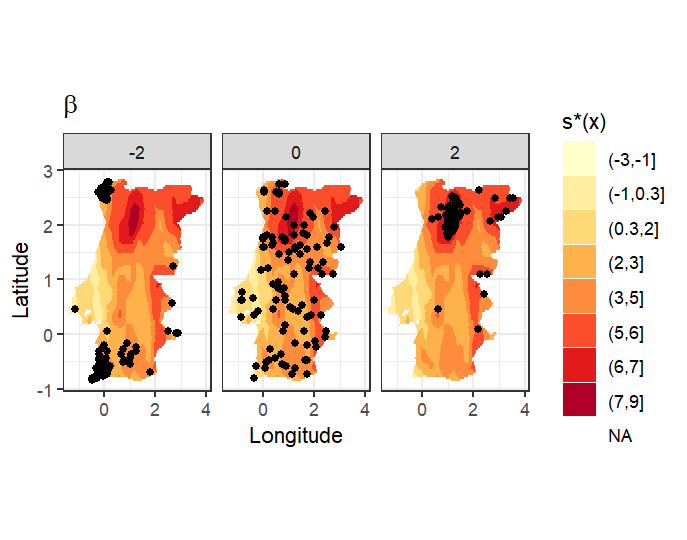}
\caption{Realization of a Mat\'{e}rn field  with $\lambda=1$,  $\sigma^2=2.5$, range $r=1.5$ and mean $\mu=4$ (left) or mean $\mu=dist(x)$ (right) and preferential and non preferential simulated data ($\beta \in \{-2,0,2\}$).}
	\label{NatarioEtal:fig1}
\end{figure}

The study 
considered a non-regular domain of interest $\cal{D}$, and simulated true realizations $s^*(x)$ of a Mat\'{e}rn field with $\lambda=1$,  $\sigma^2=2.5$ and range $r=1.5$. The true field mean was allowed to be either $\mu=4$ or given by a spatial covariate representing the Euclidean distance to coast (respectively, in the left and bottom sides of the considered domain), $dist(x)$. Data simulation followed and, for the same field realization, $n=100$ points were sampled under different preferentiability scenarios, ranging from strong positive to strong negative preferentiability, as well as no preferentiability. Samples were drawn from a Gaussian distribution with mean
$\beta s^*(x)$, where $\beta=2$ and $\beta=-2$ represent the most intense preferentiability settings, and $\beta=0$ corresponds to the non-preferential case. Intermediate scenarios were generated using $\beta \in \{1.5, 1.0, 0.5, -0.5, -1.0, -1.5 \}$.  A nugget variance of $\tau^2=0.2$ was assumed.  
This process was repeated for each spatial process considered, simulating 50 different realizations of the corresponding spatial field and data sets for each different level of preferential sampling (determined by the $\beta$ value). See Figure \ref{NatarioEtal:fig1} for a realization of each of the spatial processes described and corresponding simulated data for three different values of $\beta$.

\begin{figure}
\centering
\includegraphics[width=0.95\linewidth, trim=-1cm 2.5cm -0.5cm 2cm, clip]{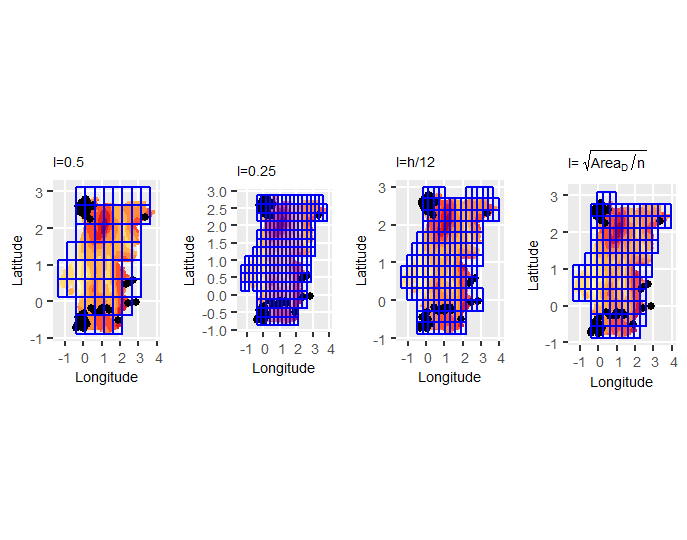}  
\caption{Grids considered in the tests for the simulated data set corresponding to $\beta=-2$. Grid side sizes from left to right: $l=0.5$, $l=0.25$, $l=h/12$, $l=\sqrt{\mbox{Area}/n}$.  }
	\label{NatarioEtal:fig2}
\end{figure}

Four grid resolutions were considered, varying the side length of the square cells to assess the sensitivity of the test to spatial aggregation, varying the square grid side size $l$: $l={h \over 12}$, where $h$ is the maximum distance between all the $n$ sampled points; the square root of the average area per point, representing the side length of a square with area given by total domain area divided by the number of sampled points, $l=\sqrt{\mbox{Area}/n}$; $l=0.5$; $l=0.25$. The latter two values were chosen to represent 5\% and 2.5\%, respectively, of the maximum straight-line distance between the east-west and north-south boundaries of the domain.  We should note that if the grid size is too small, such that each cell contains only one data point, that naturally compromises the test results.  Figure \ref{NatarioEtal:fig2} displays the grids corresponding to the first data set displayed in Figure \ref{NatarioEtal:fig1}.

The grid side size $l={h \over 12}$, in the examples of Figure \ref{NatarioEtal:fig1}, varied in $\left\{0.396, 0.349, 0.398,  0.298, 0.349, 0.339\right \}$, from left to right. In the whole simulation study, over all the different simulation conditions, the values of $l={h \over 12}$ varied between  0.130 and  0.466, with a mean value of 0.379, for the case of the true simulated field mean of $4$, and varied between 0.109 and 0.457, with a mean value of 0.359 for the case of the true field mean given by distance to coast $dist(x)$. As for the values of $l=\sqrt{\mbox{Area}_{\mathcal{D}}/n}$, this was constant equal to 0.332.


\clearpage

\begin{figure}
	\centerline{
\includegraphics[width=10cm]{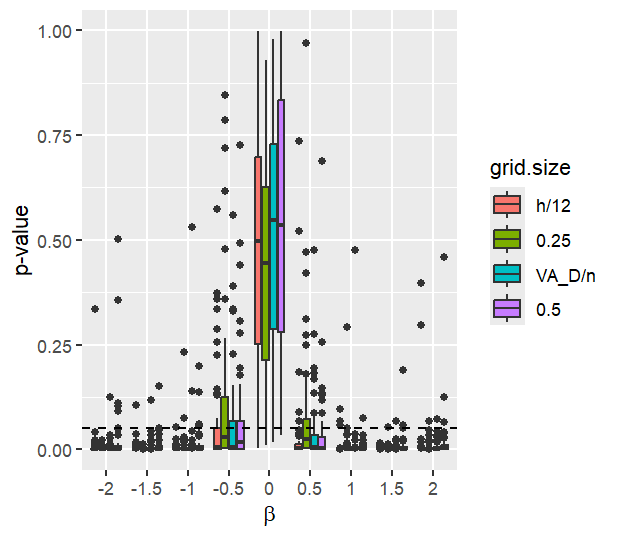}}
\caption{Boxplots of the p-values for the 50 replicas of the MLC test for each value of $\beta$ considered in the data simulation and each value of grid size considered in the test, for the case where the true field mean was $4$. The dashed line represents the 0.05 threshold.  }
	\label{NatarioEtal:fig3}
\end{figure}

\begin{figure}
	\centerline{
\includegraphics[width=10cm]{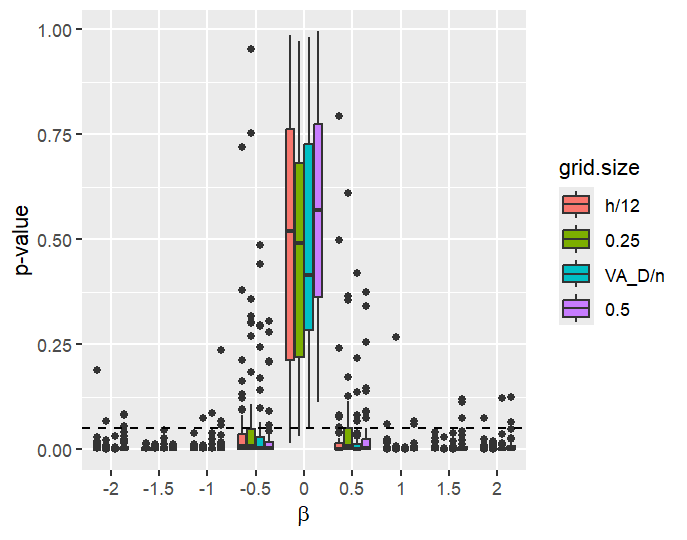}}
\caption{Boxplots of the p-values for the 50 replicas of the MLC test for each value of $\beta$ considered in the data simulation and each value of grid size considered in the test, for the case where the true field mean was distance to coast $dist(x)$. The dashed line represents the 0.05 threshold.   }
	\label{NatarioEtal:fig4}
\end{figure}

Considering first the MLC test in its frequentist approach, Figure \ref{NatarioEtal:fig3} and Table \ref{natariocarvalho:table1} summarize the test results for the case where the true field mean was $4$, for the different values of $\beta$ used in the simulation of the data and for the different grid sizes detailed before. Figure \ref{NatarioEtal:fig4} and Table \ref{natariocarvalho:table2} summarize the test results for the case where the true field mean was distance to coast $dist(x)$. From these, it is possible to see that for both cases, the test only consistently accepted the non-preferential sampling hypothesis for the non preferential simulated data set ($\beta=0$), at a 5\% significance level. For the case of $\beta=0.5$ or $\beta=-0.5$, which might be considered a a relatively weak scenario scenario of preferential sampling, the test accepted the non-preferential hypothesis for 10-40\% of the times, at a 5\% significance level. These percentage diminished for 10-25\% in the second case, of mean simulated field given by distance to coast $dist(x)$.

The test performance does not seem to vary greatly with grid size. For the first case of the true simulated field mean of $4$, the best results are more associated with the grid side size of $l={h \over 12}$, quite close to that of $l=\sqrt{\mbox{Area}_{\mathcal{D}}/n}$. In the second case the best results are more associated with the grid side size of $l=\sqrt{\mbox{Area}_{\mathcal{D}}/n}$, very close however to that of $l={h \over 12}$.

\renewcommand{\arraystretch}{1.1}

\begin{center}
\begin{table}[ht]
\centering
\caption{Proportion of replicas that confirmed preferential sampling on the MLC test (p-value $<~0.05$ ), for different values of $\beta$ used to generate data and different values of grid sizes for the test, and for the case where the true field mean was $4$.}
\begin{tabular}{ccccc}
  \hline
  & \multicolumn{4}{c}{Grid size} \\ \cline{2-5}
$\beta$ & 0.5 & 0.25 & $h \over 12$ & $\sqrt{\mbox{Area}_{\mathcal{D}}\over n}$ \\
\hline
-2 & 0.90 & 1.00 & 0.98 & 0.98 \\
  -1.5 & 0.96 & 1.00 & 0.98 & 0.98 \\
  -1 & 0.92 & 0.96 & 0.98 & 0.96 \\
  -0.5 & 0.68 & 0.60 & 0.74 & 0.70 \\
  0 & 0.04 & 0.08 & 0.04 & 0.04 \\
  0.5 & 0.78 & 0.66 & 0.90 & 0.80 \\
  1 & 0.98 & 0.96 & 0.94 & 0.98 \\
  1.5 & 0.94 & 0.98 & 1.00 & 0.98 \\
  2 & 0.94 & 0.98 & 0.96 & 0.98 \\   \hline
\end{tabular}
\label{natariocarvalho:table1}
\end{table}
\end{center}

\renewcommand{\arraystretch}{1.1}

\begin{center}
\begin{table}[ht]
\centering
\caption{Proportion of replicas that confirmed  preferential sampling on the MLC test (p-value $<~0.05$ ), for different values of $\beta$ used to generate data and different values of grid sizes for the test, and for the case where the true field mean  was distance to coast $dist(x)$.}
\begin{tabular}{ccccc}
  \hline
  & \multicolumn{4}{c}{Grid size} \\ \cline{2-5}
$\beta$ & 0.5 & 0.25 & $h \over 12$ & $\sqrt{\mbox{Area}_{\mathcal{D}}\over n}$ \\
\hline
-2 & 0.94 & 0.98 & 0.98 & 1.00 \\
  -1.5 & 1.00 & 1.00 & 1.00 & 1.00 \\
  -1 & 0.94 & 0.98 & 1.00 & 0.98 \\
  -0.5 & 0.88 & 0.76 & 0.78 & 0.82 \\
  0 & 0.00 & 0.02 & 0.06 & 0.00 \\
  0.5 & 0.82 & 0.74 & 0.86 & 0.90 \\
  1 & 0.96 & 0.98 & 0.98 & 1.00 \\
  1.5 & 0.94 & 1.00 & 1.00 & 1.00 \\
  2 & 0.96 & 1.00 & 0.98 & 0.98 \\    \hline
\end{tabular}
\label{natariocarvalho:table2}
\end{table}
\end{center}

\begin{figure}
	\centerline{
\includegraphics[width=0.49\linewidth]{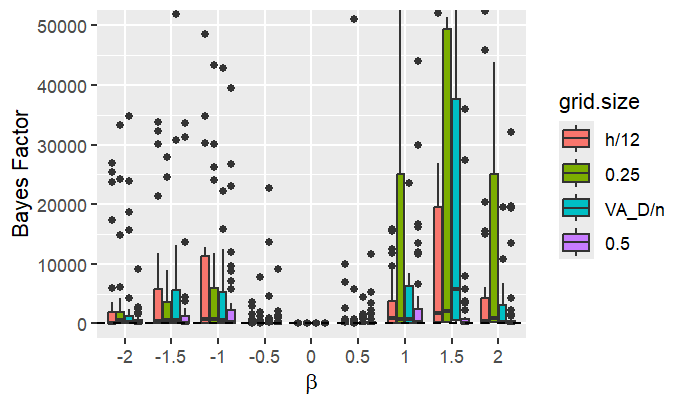}
\includegraphics[width=0.49\linewidth]{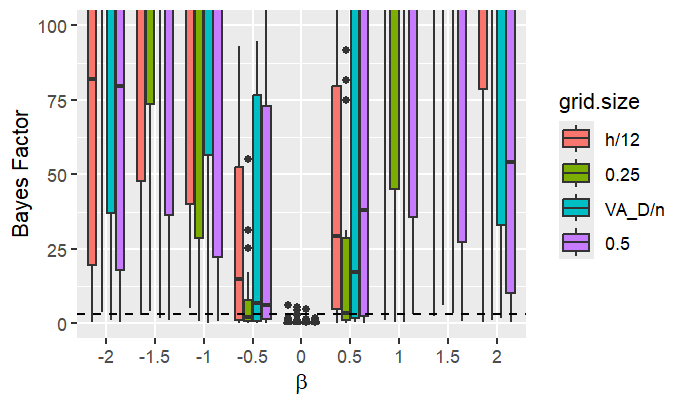}}
\caption{Boxplots of the Bayes factor $B_{10}$ for the 50 replicas of the MLC test for each value of $\beta$ considered in the data simulation and each value of grid size considered in the test, for the case where the true field mean was $4$. The threshold of 3 for the Bayes factor $B_{10}$ is depicted for reference on  positive evidence for preferential. The figure on the right zooms in the figure on the left.}
	\label{NatarioEtal:fig5}
\end{figure}

\begin{figure}
	\centerline{
\includegraphics[width=0.49\linewidth]{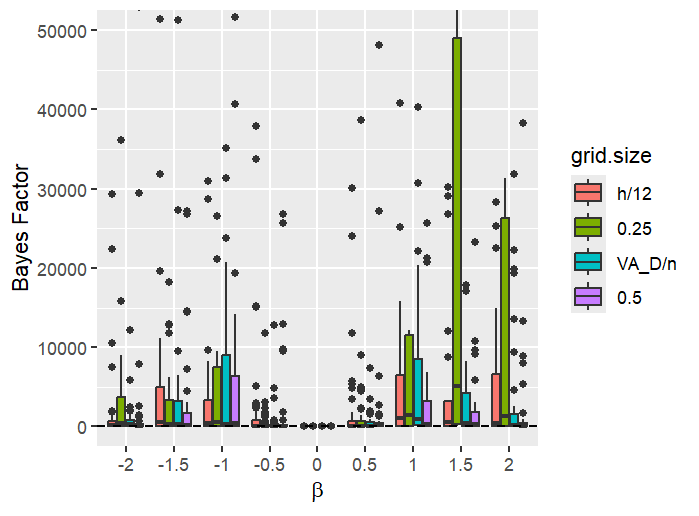}
\includegraphics[width=0.49\linewidth]{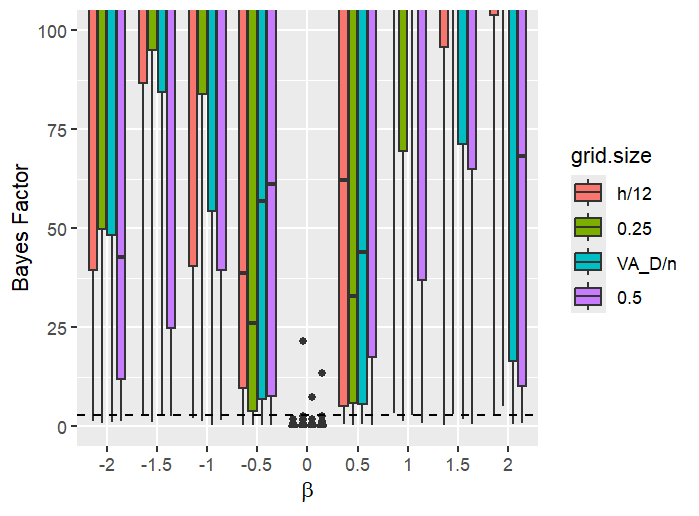}}
\caption{Boxplots of the Bayes factor $B_{10}$ for the 50 replicas of the MLC test for each value of $\beta$ considered in the data simulation and each value of grid size considered in the test,  for the case where the true field mean was distance to coast $dist(x)$. The threshold of 3 for the Bayes factor $B_{10}$ is depicted for reference on  positive evidence for preferential. The figure on the right zooms in the figure on the left.  }
	\label{NatarioEtal:fig6}
\end{figure}

The following step on this simulation study was to run the MLC test in its Bayesian formulation. Figure \ref{NatarioEtal:fig5} and Table \ref{natariocarvalho:table3} summarize the test results for the case corresponding to a true field mean of  $4$, for different values of $\beta$ and different grid sizes for the MLC test. Figure \ref{NatarioEtal:fig6} and Table \ref{natariocarvalho:table4} summarize the test results for the case of true field mean given by the distance to coast $dist(x)$. For both cases, the test consistently indicated no preferential sampling for the non preferential simulated data set ($\beta=0$), corresponding to a Bayes factor $B_{10}$ greater than 3 (positive evidence). For the case of $\beta=0.5$ or $\beta=-0.5$, the test accepted the non-preferential hypothesis for about 20-30\% of the times for the  case of a mean simulated field of $4$, and accepted the non-preferential hypothesis for about 20\% of the times for the  case of mean simulated field given by distance to coast $dist(x)$. Again, the test performance does not seem to vary greatly with grid size.

The simulation study was additionally repeated for different sample sizes $n$, that varied further in $\{50, 250, 500 \}$, summary of results in Appendix \ref{NETAL:ApA}. For $n=250$ and $n=500$, the test results identified preferentiability 100\% of the times for $\beta \in \{-2, -1.5, -1.0, 1.0, 1.5, 2,0\}$, more than 98\% of the times for $\beta \in \{-0.5, 0.5\}$ and generally less than 10\% for $\beta=0$. For $n=50$, preferentiability was generally identified around 80\% of the times for $\beta \in \{-2, -1.5, -1.0, 1.0, 1.5, 2,0\}$, around $35\%$ of the times for $\beta \in \{-0.5, 0.5\}$ and less than 5\% for $\beta=0$. These finding should be taken into consideration when applying this test, in conjunction with the test results, as for small sample sizes the test tends to favor the non-preferentiability hypothesis whereas for large sample sizes it is the opposite case.

The simulation results demonstrate that the MLC test is a reliable and computationally efficient tool for detecting preferential sampling across a wide range of scenarios. Its robustness to grid resolution and mean structure, combined with its simplicity, makes it suitable for routine use in applied geostatistics.

\renewcommand{\arraystretch}{1.1}

\begin{center}
\begin{table}[ht]
\centering
\caption{Proportion of replicas that confirmed preferential sampling  on the MLC test ($B_{10}~>~3$ ), for different values of $\beta$ used to generate data and different values of grid sizes for the test, and for the case where the true field mean was $4$.}
\begin{tabular}{ccccc}
  \hline
  & \multicolumn{4}{c}{Grid size} \\ \cline{2-5}
$\beta$ & 0.5 & 0.25 & $h \over 12$ & $\sqrt{\mbox{Area}_{\mathcal{D}}\over n}$ \\
\hline
-2 & 0.86 & 1.00 & 0.94 & 0.96 \\
  -1.5 & 0.94 & 1.00 & 0.98 & 0.94 \\
  -1 & 0.88 & 0.92 & 1.00 & 0.92 \\
  -0.5 & 0.58 & 0.44 & 0.66 & 0.62 \\
  0 & 0.00 & 0.06 & 0.02 & 0.02 \\
  0.5 & 0.72 & 0.52 & 0.86 & 0.66 \\
  1 & 1.00 & 0.94 & 0.94 & 0.98 \\
  1.5 & 0.96 & 1.00 & 0.98 & 1.00 \\
  2 & 0.92 & 0.96 & 0.96 & 0.94 \\ \hline
\end{tabular}
\label{natariocarvalho:table3}
\end{table}
\end{center}

\clearpage

\renewcommand{\arraystretch}{1.1}

\begin{center}
\begin{table}[ht]
\centering
\caption{Proportion of replicas that confirmed preferential sampling  on the MLC test ($B_{10}~>~3$ ), for different values of $\beta$ used to generate data and different values of grid sizes for the test, and for the case where the true field mean  was distance to coast $dist(x)$.}
\begin{tabular}{ccccc}
  \hline
  & \multicolumn{4}{c}{Grid size} \\ \cline{2-5}
$\beta$ & 0.5 & 0.25 & $h \over 12$ & $\sqrt{\mbox{Area}_{\mathcal{D}}\over n}$ \\
\hline
-2 & 0.90 & 0.96 & 0.96 & 0.96 \\
  -1.5 & 0.98 & 0.94 & 0.98 & 0.98 \\
  -1 & 0.98 & 0.94 & 0.98 & 0.96 \\
  -0.5 & 0.84 & 0.76 & 0.82 & 0.82 \\
  0 & 0.02 & 0.02 & 0.00 & 0.02 \\
  0.5 & 0.88 & 0.80 & 0.80 & 0.82 \\
  1 & 0.96 & 0.98 & 1.00 & 0.98 \\
  1.5 & 0.94 & 1.00 & 0.96 & 0.98 \\
  2 & 0.84 & 1.00 & 0.98 & 0.94 \\ \hline
\end{tabular}
\label{natariocarvalho:table4}
\end{table}
\end{center}

\clearpage

\section{Real Data Study}\label{NETAL:sec4}

In this section, two real geostatistical data sets were evaluated for their sampling preferentiability.

\begin{figure}
\centering
\includegraphics[width=0.3\textwidth,height=4cm,trim=1cm 2.5cm 0.5cm 2cm, clip]{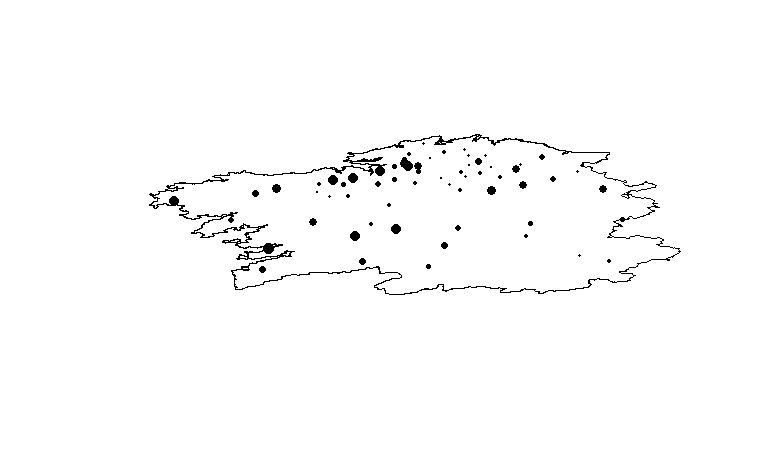}
\hspace{2cm}
\includegraphics[width=0.3\textwidth,height=4cm,trim=1cm 2.5cm 0.5cm 2cm, clip]{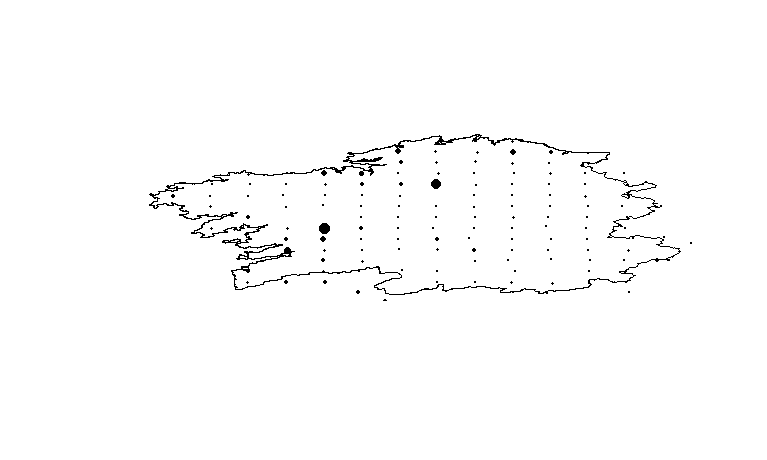}
\caption{Lead concentrations measured at the sampled locations in Galicia (concentration in micrograms per gram of dry moss) measured in the marked locations, for the year of 1997 (left) and for the year of 2000 (right). The size of the mark is proportional to the lead concentration}
	\label{NatarioEtal:fig7}
\end{figure}

The first example refers to data on lead concentrations (in micrograms per gram of dry moss) in moss samples collected in Galicia, taken in two different occasions, 1997 and 2000. These data are available from the discontinued R-package PrevMap and, for the year of 1997, have been indicated as corresponding to preferential sampled data, \cite{DiggleEtal2010} and \cite{Watson2021}. Figure \ref{NatarioEtal:fig7} depicts the $n=63$ locations of the data for 1997 and the $n=132$ locations of the data for 2000, with the size of the mark being proportional to the lead concentration. Note that there is a larger number of observations towards the north of the region for the 1997 case. For the year of 2000, the sampling designed has been changed, no longer being preferential.


\begin{figure}
\centering
\includegraphics[width=0.49\linewidth, trim=3cm 2.5cm 3cm 2cm, clip]{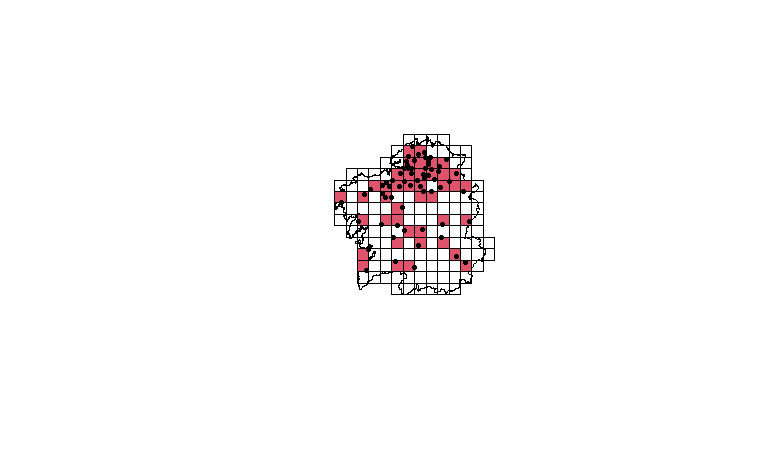}  
\includegraphics[width=0.49\linewidth, trim=3cm 2.5cm 3cm 2cm, clip]{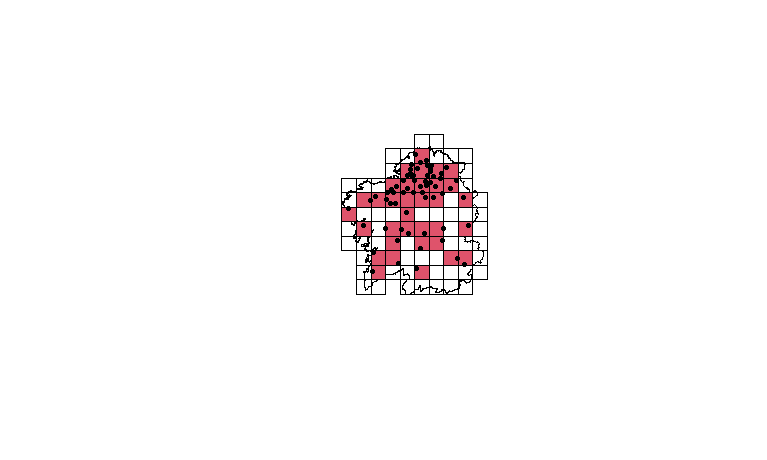}
\caption{Grids used for the MLC test for the 1997 data, for a grid size of $l = {h \over 12}$ (left) and a grid size of $l=\sqrt{\mbox{Area}_{\mathcal{D}}/n}$ (right). The red cells are those that contribute to the test}
	\label{NatarioEtal:fig8}
\end{figure}

\begin{figure}
	\centerline{
\includegraphics[width=0.49\linewidth, trim=3cm 2.5cm 3cm 2cm, clip]{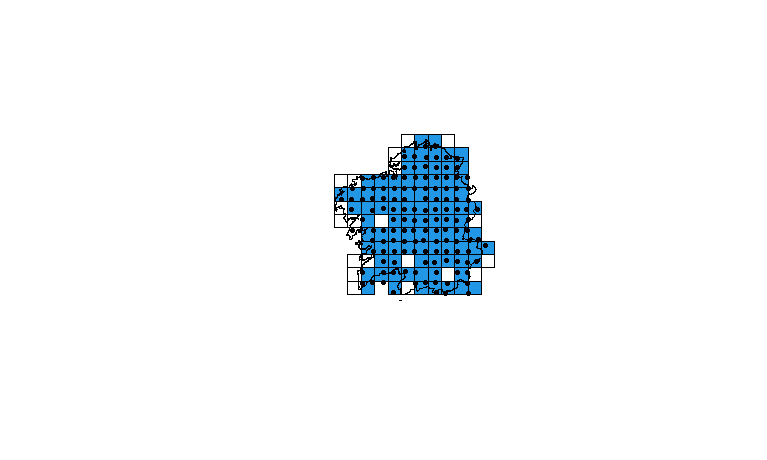}
\includegraphics[width=0.49\linewidth, trim=3cm 2.5cm 3cm 2cm, clip]{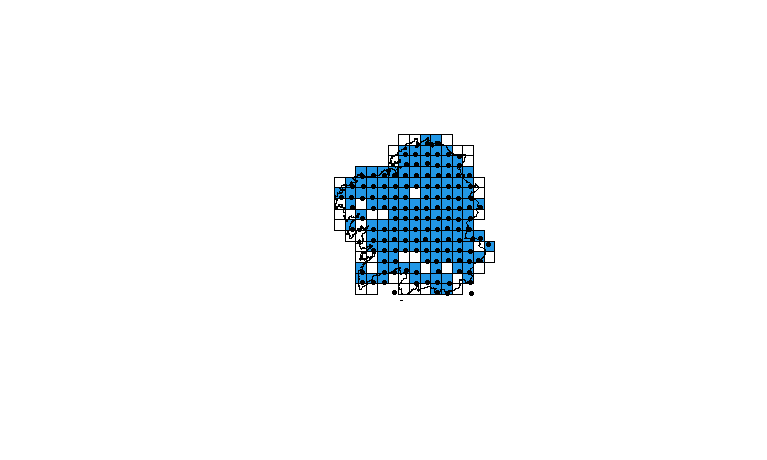}}
\caption{Grids used for the MLC test for the 2000 data, for a grid size of $l = {h \over 12}$ (left) and a grid size of $l=\sqrt{\mbox{Area}_{\mathcal{D}}/n}$ (right). The blue cells are the ones that contribute for the test}
	\label{NatarioEtal:fig9}
\end{figure}

For each data set, the frequentist MLC test was performed, for two different grid size values: $l = {h \over 12}$, with $h$ being the maximum distance between all the $n$ points in each data set ($l=16030.93$m for 1997 and $l=19037.57$m for 2000) and; $l=\sqrt{\mbox{Area}_{\mathcal{D}}/n}$, the square root of the average area per point ($l=21662.51$m for 1997 and $l=14965.53$m for 2000). Figure \ref{NatarioEtal:fig8} represents the different grids for the test of the 1997 data, standing out the red cells as the ones that contributed for the test. Figure \ref{NatarioEtal:fig9} represents the different grids for the test of the 2000 data, standing out the blue cells that were the contributors for the test. Table \ref{natariocarvalho:table5} presents the corresponding MLC frequentist observed test statistics and p-values, from which we conclude that preferential sampling is detected in 1997 but not in 2000, confirming previous conclusion. Note that these conclusions are strengthened by the respective sample sizes of each year, as they go against the favored conclusion associated with their respective sample sizes.

\renewcommand{\arraystretch}{1.2}
\begin{center}
\begin{table}[ht]
\centering
\caption{MLC frequentist test statistics ($r_S$) and p-value for accessing preferentiability in Galicia lead concentration datasets, for 1997 and 2000.}
\begin{tabular}{c|cc|cc}
  \hline
  & \multicolumn{4}{c}{Year} \\ \cline{2-5}
Grid size & \multicolumn{2}{c|}{1997} & \multicolumn{2}{c}{2000}   \\ \cline{2-3} \cline{4-5}
& $r_S$ & p-value  &  $r_S$ & p-value \\
\hline
$h \over 12$ & -0.4095 & 0.0071 & -0.1246 &  0.2503 \\
  $\sqrt{\mbox{Area}_{\mathcal{D}}\over n}$ & -0.3267 & 0.0593 & 0.0397 & 0.6565 \\
 \hline
\end{tabular}
\label{natariocarvalho:table5}
\end{table}
\end{center}

Next, for each data set, the bayesian MLC test was performed, for the same two grids as before. Table \ref{natariocarvalho:table6} gives the results for the corresponding MLC bayesian test, the estimated Spearman correlation ($\hat{\rho}_S$) and the bayes factor, $B_{10}$, indicating that preferential sampling is detected in 1997 but not in 2000.

\renewcommand{\arraystretch}{1.2}
\begin{center}
\begin{table}[ht]
\centering
\caption{Bayesian MLC test results (estimated Spearman correlation $\hat{\rho}_S$ and Bayes factor) for accessing preferentiability in Galicia lead concentration datasets, for 1997 and 2000.}
\begin{tabular}{c|cc|cc}
  \hline
  & \multicolumn{4}{c}{Year} \\ \cline{2-5}
Grid size & \multicolumn{2}{c|}{1997} & \multicolumn{2}{c}{2000}   \\ \cline{2-3} \cline{4-5}
& $\hat{\rho}_S$ & $B_{10}$  &  $\hat{\rho}_S$ & $B_{10}$ \\
\hline
$h \over 12$ & -0.4323 & 5.5256 & -0.1435 &  0.3063 \\
  $\sqrt{\mbox{Area}_{\mathcal{D}}\over n}$ & -0.3222 & 1.1671 & 0.1150 & 0.4161 \\
 \hline
\end{tabular}
\label{natariocarvalho:table6}
\end{table}
\end{center}

\begin{figure}
	\centerline{
\includegraphics[width=0.6\linewidth, height=4cm, trim=1cm 2.5cm 1cm 2cm, clip]{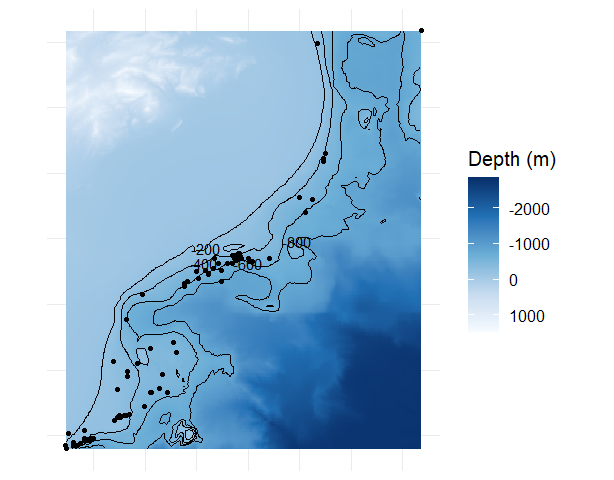}}  
\caption{Blue and red-shrimp data, representing the fishery locations and bathymetric information}
	\label{NatarioEtal:fig10}
\end{figure}

\begin{figure}
	\centerline{
\includegraphics[width=0.38\linewidth, height=3.8cm, trim=1cm 2.5cm 1cm 2cm, clip]{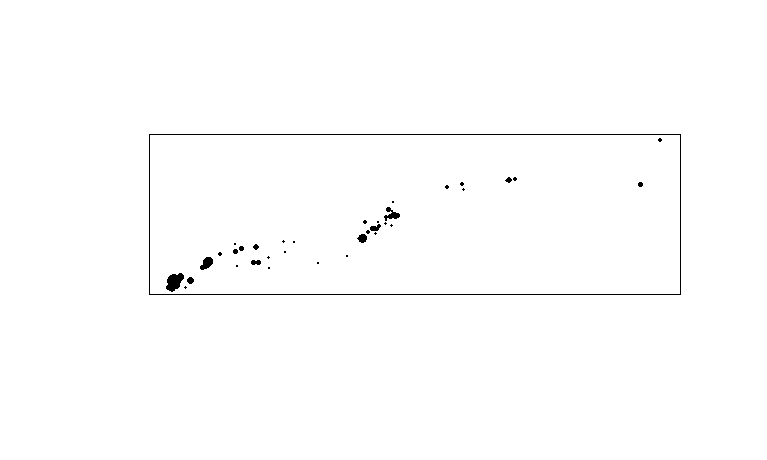}  
\includegraphics[width=0.38\linewidth, height=3.8cm, trim=1cm 2.5cm 1cm 2cm, clip]{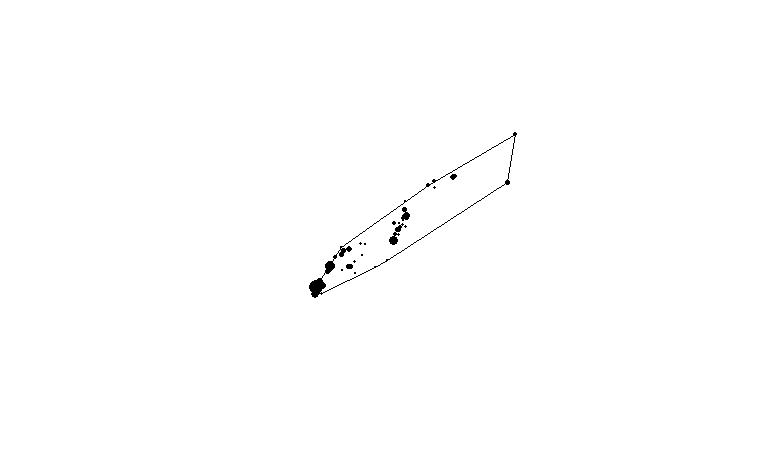}}
\caption{Blue and red-shrimp data, representing the quantity of fish caught (Kg) measured in the fishery locations, for a rectangle area (left) and a more restricted area containing the observations (right). The size of the mark is proportional the caught quantity}
	\label{NatarioEtal:fig11}
\end{figure}

The second example refers to data on the caught locations and weight (Kg) of blue and red shrimp (\textit{Aristeus antennatus, Risso 1816}), by onboard observers of a number of fishing boats in the Gulf of Alicante (Spain) from 2009 to 2012, \cite{Pennino2019}, kindly made available by M. G. Pennino. Figure \ref{NatarioEtal:fig10} depicts the $n=86$ locations and bathymetric information, while Figure \ref{NatarioEtal:fig11} depicts the locations with the size proportional to the quantity of fish caught.

In order to perform the MLC test, the domain region needed to be established. Given the data location distribution, one might be tempted to frame the data in a rectangle, as in the left panel of Figure \ref{NatarioEtal:fig11} (rectangle area), which ends up to have a large part of the domain with no data at all, corresponding to places that do not have the appropriate depth conditions to catch blue and red shrimp - as seen in Figure \ref{NatarioEtal:fig10}. Alternatively, one may consider the domain restricted to the area where blue and red shrimp were looked for and caught, corresponding to ideal bathymetric conditions, as in the right panel of Figure \ref{NatarioEtal:fig11} (restricted area). This nuance may naturally have an impact on the MLC test results, specially when the grid size is chosen as a function of the domain area.

\begin{figure}
\centering
\includegraphics[width=0.35\linewidth, trim=3cm 2.5cm 3cm 2cm, clip]{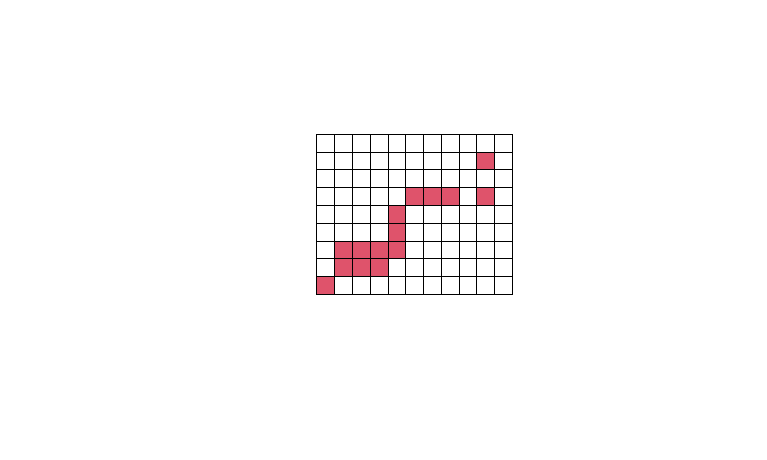}  
\includegraphics[width=0.35\linewidth, trim=3cm 2.5cm 3cm 2cm, clip]{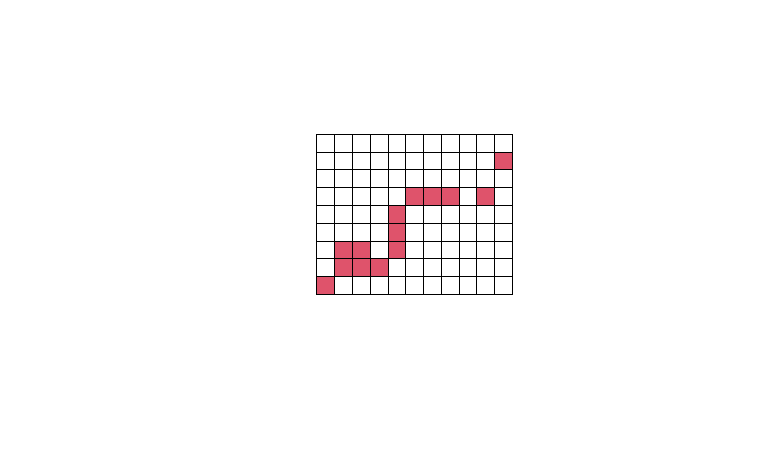}
\caption{Grids used for the MLC test for the blue and red shrimp data in the rectangle domain, for a grid size of $l = {h \over 12}$ (left) and a grid size of $l=\sqrt{\mbox{Area}_{\mathcal{D}}/n}$ (right). The red cells are the ones that contribute for the test.}
	\label{NatarioEtal:fig12}
\end{figure}

\begin{figure}
\centering
\includegraphics[width=0.35\linewidth, trim=3cm 2.5cm 3cm 2cm, clip]{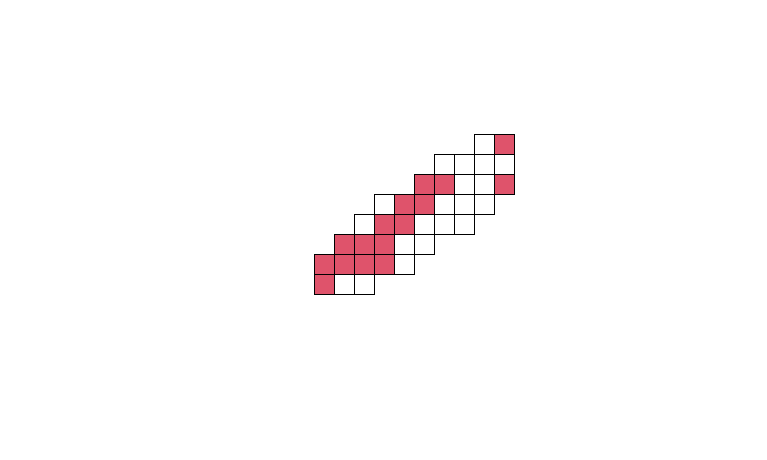}  
\includegraphics[width=0.35\linewidth, trim=3cm 2.5cm 3cm 2cm, clip]{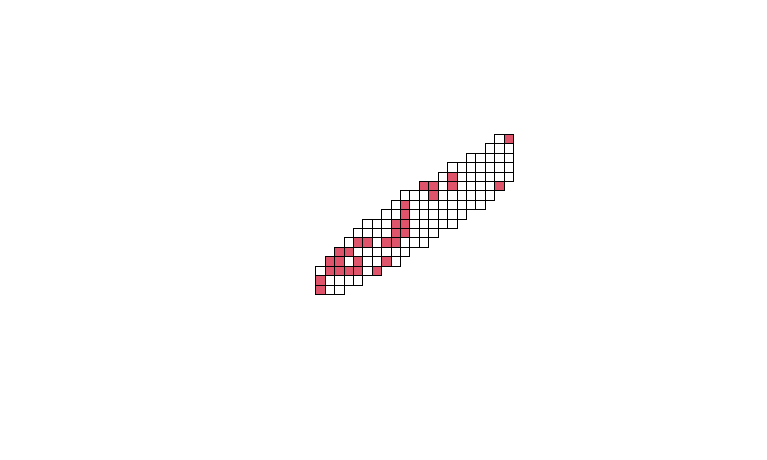}
\caption{Grids used for the MLC test for the blue and red shrimp data in the restricted domain, for a grid size of $l = {h \over 12}$ (left) and a grid size of $l=\sqrt{Area_D/n}$ (right). The red cells are the ones that contribute for the test.}
	\label{NatarioEtal:fig13}
\end{figure}

For each different domain, the frequentist MLC test was performed, for the different grid size values: $l = {h \over 12}$, with $h$ being the maximum distance between all the $n$ points in each data set ($l=15494.65$m for both domains) and; $l=\sqrt{Area_D/n}$, the square root of the average area per point ($l=15140.48$m for the rectangle area and $l=7078.62$m for the restricted area). Figure \ref{NatarioEtal:fig12} represents the different grids for the test with the rectangle domain, while Figure \ref{NatarioEtal:fig13} represents the same for the test considering the restricted domain, standing out the red cells as the ones that contribute for each test. Table \ref{natariocarvalho:table7} presents the corresponding MLC frequentist test statistics and p-value, from which we conclude that preferential sampling is detected when considering the restricted area, considering a significance level of $5\%$, confirming previous conclusions \cite{Pennino2019}, but not in the case of the rectangle domain, which actually do not represent the effective area of interest for the application. It should be noted that the not so large sample size would not favour the preferentiability hypothesis, that was accepted nevertheless.

\renewcommand{\arraystretch}{1.2}
\begin{center}
\begin{table}[ht]
\centering
\caption{MLC frequentist test statistics ($r_S$) and p-value for accessing preferentiability in the blue and red-shrimp dataset, for the rectangle and the restricted domain.}
\begin{tabular}{c|cc|cc}
  \hline
  & \multicolumn{4}{c}{Area} \\ \cline{2-5}
Grid size & \multicolumn{2}{c|}{Rectangle} & \multicolumn{2}{c}{Restricted}   \\ \cline{2-3} \cline{4-5}
& $r_S$ & p-value  & $r_S$ & p-value \\
\hline
$h \over 12$ & 0.0416 & 0.8829 & 0.4990 & 0.0491  \\
  $\sqrt{\mbox{Area}_{\mathcal{D}}\over n}$ & -0.0513 & 0.8618 & 0.4619 & 0.0102 \\
 \hline
\end{tabular}
\label{natariocarvalho:table7}
\end{table}
\end{center}

Finally, for the blue red shrimp data set, the bayesian MLC test was performed, for the two different domains and considering the two different grids sizes. Table \ref{natariocarvalho:table8} gives the results for the corresponding MLC bayesian test, the estimated Spearman correlation and the bayes factor, $B_{10}$, indicating that preferential sampling is detected only in the restricted domain case, more strongly for the grid sized based on the area.

\renewcommand{\arraystretch}{1.2}
\begin{center}
\begin{table}[ht]
\centering
\caption{Bayesian MLC test results (estimated Spearman correlation $\hat{\rho}_S$ and Bayes factor) for accessing preferentiability in the blue and red-shrimp dataset, for the rectangle and the restricted domain.}
\begin{tabular}{c|cc|cc}
  \hline
  & \multicolumn{4}{c}{Area} \\ \cline{2-5}
Grid size & \multicolumn{2}{c|}{Rectangle} & \multicolumn{2}{c}{Restricted}   \\ \cline{2-3} \cline{4-5}
& $\hat{\rho}_S$ & $B_{10}$  &  $\hat{\rho}_S$ & $B_{10}$ \\
\hline
$h \over 12$ & 0.0434 & 0.3013 & 0.4293  & 1.6919  \\
  $\sqrt{\mbox{Area}_{\mathcal{D}}\over n}$ & -0.0158 & 0.2930 & 0.4244 & 4.3495 \\
 \hline
\end{tabular}
\label{natariocarvalho:table8}
\end{table}
\end{center}


\section{Discussion}\label{NETAL:sec5}

This study demonstrates that the MLC test is a simple, robust, and broadly applicable tool for detecting preferential sampling in geostatistical settings, by viewing this as a problem of testing stochastic independency between sampling locations where to measure a spatial phenomenon ruled by a stochastic process and those same measurements. In contrast to several existing approaches, the proposed method does not rely on restrictive assumptions such as stationarity or isotropy and can be applied to both continuous and discrete marks, not making distributional assumptions regarding the measured stochastic process. Additionally, the MLC test implementation does not require fitting complex models to either the measurement or the underlying spatial processes, making it particularly accessible to practitioners with limited computational or modelling skills, possible to be implemented with standard statistical tools. Also, although it might work better for larger samples sizes, that is not a requirement.

The simulation study provides encouraging evidence regarding the performance of the proposed methodology. Both grid specifications considered, namely $l={h \over 12}$, with $h$ being the maximum distance between all sampled points, and $l=\sqrt{\mbox{Area}_{\mathcal{D}}/n}$, produced very similar results and frequently corresponded to the best-performing configurations in both frequentist and Bayesian implementations. Importantly, the proposed approaches were generally successful in identifying preferential sampling whenever it was present. Even under relatively weak preferentiality scenarios ($\beta=0.5$ or $\beta=-0.5$), correct identification rates exceeded 75\% in most cases, suggesting that the method remains informative even when preferentiality is only moderately expressed. It should be noted, however, that there is an expected influence in test results regarding the sample size, with large sample sizes favouring the identification of preferential sampling and smaller sample sizes having more difficulty in identifying such scenarios. The results should always be seen in the light os the corresponding sample sizes.

The comparison between the frequentist and Bayesian implementations demonstrated a similar ability to detect preferences, although to the Bayesian version corresponds a higher computational cost due to the MCMC simulations involved. Even so, the computation times proved feasible in the scenarios considered, approximately two minutes per dataset. This additional computational load can be justified when we want to quantify the posterior uncertainty or make probabilistic statements about the dependence.

The real-data applications further support the practical utility of the MLC test. In all analysed cases, the proposed method identified preferentiality whenever previous studies had also reported its presence, lending empirical credibility to the approach. However, these applications also highlight an important point: correct specification of the study domain is crucial, since inappropriate domain definition may confound the dependence structure and lead to misleading conclusions regarding preferentiality.

Some limitations and opportunities for methodological refinement should also be acknowledged. Although large sample sizes are not formally required for the application of the MLC test, performance may improve as sample size increases, particularly when the dependence structure is weak or heterogeneous. Furthermore, because spatial dependence may compromise the assumption of independence among observations, the Spearman correlation that assumes independence between observations should account for this relationship, \citet{DuncanEtal2014}. Failing to account spatial autocorrelation may result in an incorrect assessment of the corresponding test p-values, as the number of degrees of freedom of the distribution of the test statistics is inaccurately determined and should be corrected. This can be addressed by using an effective sample size correction approach, \citet{CliffordEtal1989}, although it was chosen not to do so here, leaving this to a future work.

The MLC test constitutes a practical, computationally accessible, and possible widely applicable tool for detecting preferential samples. By avoiding strong distributional assumptions and complex modeling, while maintaining good empirical performance, the proposed method offers a valuable alternative for researchers working with preferentially sampled spatial data. Taken together, these position the MLC test as a robust addition to the field of spatial analysis, representing a meaningful contribution to applied spatial statistics with clear potential for practical adoption.

\section*{Acknowledgments}
\noindent This work is funded by national funds through the FCT - Funda\c{c}\~{a}o para a Ci\^{e}ncia e a Tecnologia, I.P., under the scope of the projects UID/00297/2025 (\url{https://doi.org/10.54499/UID/00297/2025}) and UID/PRR/00297/2025 (\url{https://doi.org/10.54499/UID/PRR/00297/2025}) (Center for Mathematics and Applications). We acknowledge Prof. M. Luc\'{\i}lia Carvalho for having the seminal idea of the MLC test and we dedicate this work to her.

\section*{Data Availability and Code}
\noindent All data and code are available in the repository \url{https://github.com/inatario/MLC_test}{https://github.com/inatario/MLC\_test}.

\clearpage

\section*{Appendix}

\appendix

\section{MLC test results for sample sizes of $n=50$, $n=250$ and $n=500$}
\label{NETAL:ApA}

The simulation process was repeated for different sample sizes: $n=50$, $n=250$ and $n=500$. Table \ref{natariocarvalho:tablex} presents a summary of the values of the grid side size  $l={h \over 12}$ over the replicas, for both cases of true field mean $4$ and $dist(x)$, and values of the grid side size $l=\sqrt{\mbox{Area}_{\mathcal{D}}/n}$, constant for each sample size.

\begin{center}
\begin{table}[ht]
\centering
\caption{Summary values of the grid side size $l={h \over 12}$ over the replicas, for different sample sizes $n$, for both cases of true field mean $4$ and $dist(x)$, and values of the grid side size $l=\sqrt{\mbox{Area}_{\mathcal{D}}/n}$, constant for each sample size.}
\begin{tabular}{c|ccc|ccc|c}
  \hline
  & \multicolumn{6}{c|}{$l={h \over 12}$} & \\ \cline{2-7}
  & \multicolumn{3}{c|}{$\mu=4$} & \multicolumn{3}{c|}{$\mu=dist(x)$} & $l=\sqrt{\mbox{Area}_{\mathcal{D}}/n}$  \\ \cline{2-7}
  $n$ & Min & Mean & Max & Min & Mean & Max \\ \hline
  50 & 0.070 & 0.355 & 0.466 & 0.093 & 0.332 & 0.451 & 0.470 \\
  100 &0.130 &0.379 & 0.466 & 0.109 & 0.359 & 0.457 & 0.332\\
  250 & 0.149 & 0.407 & 0.468 & 0.140 & 0.387 & 0.466 & 0.210\\
  500 & 0.248 & 0.423 & 0.468 & 0.242 & 0.409 &0.468 & 0.148 \\ \hline
\end{tabular}
\label{natariocarvalho:tablex}
\end{table}
\end{center}

\begin{figure}
	\centerline{
\includegraphics[width=10cm]{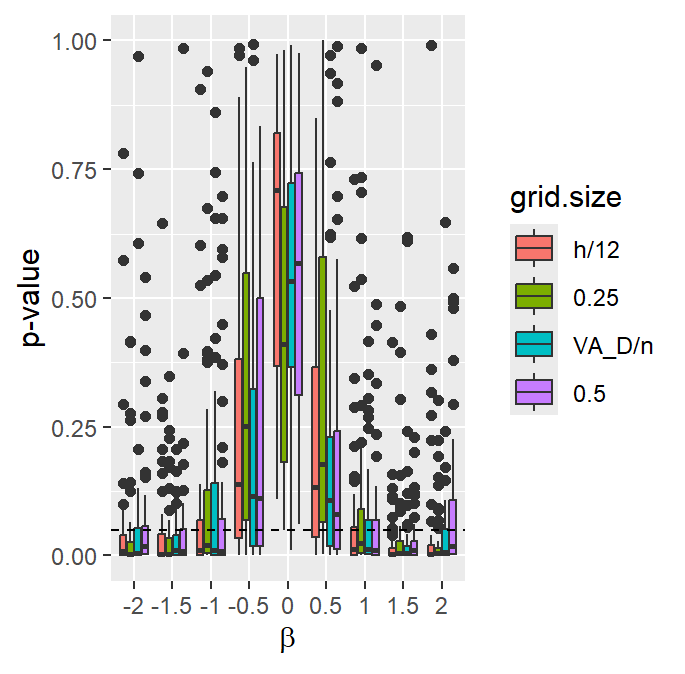}}
\caption{Boxplots of the p-values for the 50 replicas of the MLC test for each value of $\beta$ considered in the data simulation and each value of grid size considered in the test, for the case where the true field mean was $4$, $n=50$. The dashed line represents the 0.05 threshold.  }
	\label{NatarioEtal:figyy1}
\end{figure}

\begin{figure}
	\centerline{
\includegraphics[width=10cm]{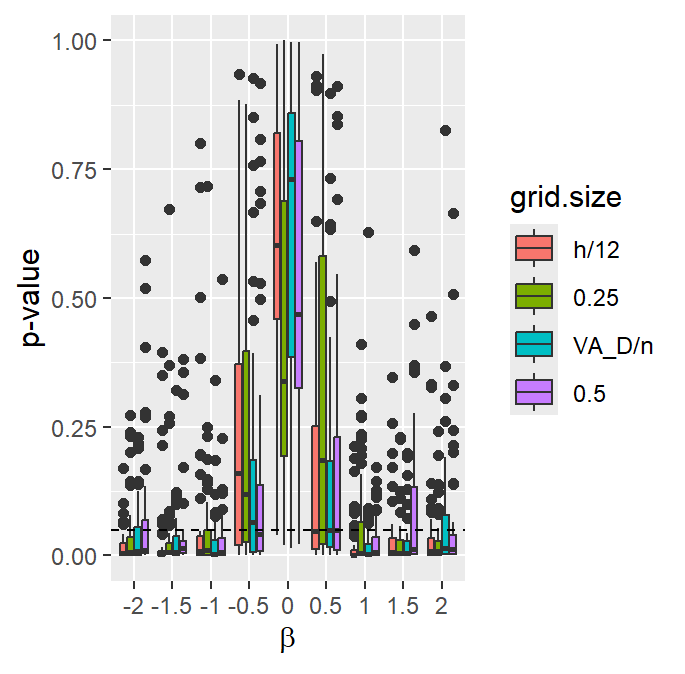}}
\caption{Boxplots of the p-values for the 50 replicas of the MLC test for each value of $\beta$ considered in the data simulation and each value of grid size considered in the test, for the case where the true field mean was distance to coast $dist(x)$, $n=50$. The dashed line represents the 0.05 threshold.   }
	\label{NatarioEtal:figyy2}
\end{figure}

\renewcommand{\arraystretch}{1.1}

\begin{center}
\begin{table}[ht]
\centering
\caption{Proportion of replicas that confirmed preferential sampling on the MCL test (p-value $<~0.05$ ), for different values of $\beta$ used to generate data and different values of grid sizes for the test, and for the case where the true field mean was $4$ and $n=50$.}
\begin{tabular}{ccccc}
  \hline
  & \multicolumn{4}{c}{Grid size} \\ \cline{2-5}
$\beta$ & 0.5 & 0.25 & $h \over 12$ & $\sqrt{\mbox{Area}_{\mathcal{D}}\over n}$ \\
\hline
-2 & 0.72 & 0.82 & 0.80 & 0.74 \\
  -1.5 & 0.74 & 0.80 & 0.76 & 0.80 \\
  -1 & 0.72 & 0.62 & 0.68 & 0.60 \\
  -0.5 & 0.40 & 0.18 & 0.32 & 0.44 \\
  0 & 0.00 & 0.00 & 0.00 & 0.06 \\
  0.5 & 0.40 & 0.18 & 0.32 & 0.40 \\
  1 & 0.62 & 0.68 & 0.74 & 0.68 \\
  1.5 & 0.82 & 0.82 & 0.84 & 0.82 \\
  2 & 0.58 & 0.80 & 0.84 & 0.74 \\   \hline
\end{tabular}
\label{natariocarvalho:tablexx1}
\end{table}
\end{center}

\renewcommand{\arraystretch}{1.1}

\vspace{-1cm}

\begin{center}
\begin{table}[ht]
\centering
\caption{Proportion of replicas that confirmed  preferential sampling on the MCL test (p-value $<~0.05$ ), for different values of $\beta$ used to generate data and different values of grid sizes for the test, and for the case where the true field mean  was distance to coast $dist(x)$ and $n=50$.}
\begin{tabular}{ccccc}
  \hline
  & \multicolumn{4}{c}{Grid size} \\ \cline{2-5}
$\beta$ & 0.5 & 0.25 & $h \over 12$ & $\sqrt{\mbox{Area}_{\mathcal{D}}\over n}$ \\
\hline
-2 & 0.74 & 0.82 & 0.88 & 0.74 \\
  -1.5 & 0.88 & 0.80 & 0.88 & 0.80 \\
  -1 & 0.78 & 0.74 & 0.82 & 0.82 \\
  -0.5 & 0.54 & 0.36 & 0.34 & 0.46 \\
  0 & 0.02 & 0.04 & 0.02 & 0.04 \\
  0.5 & 0.52 & 0.34 & 0.52 & 0.50 \\
  1 & 0.76 & 0.74 & 0.86 & 0.86 \\
  1.5 & 0.64 & 0.84 & 0.86 & 0.86 \\
  2 & 0.76 & 0.78 & 0.80 & 0.66 \\     \hline
\end{tabular}
\label{natariocarvalho:tablexx2}
\end{table}
\end{center}

\begin{figure}
	\centerline{
\includegraphics[width=10cm]{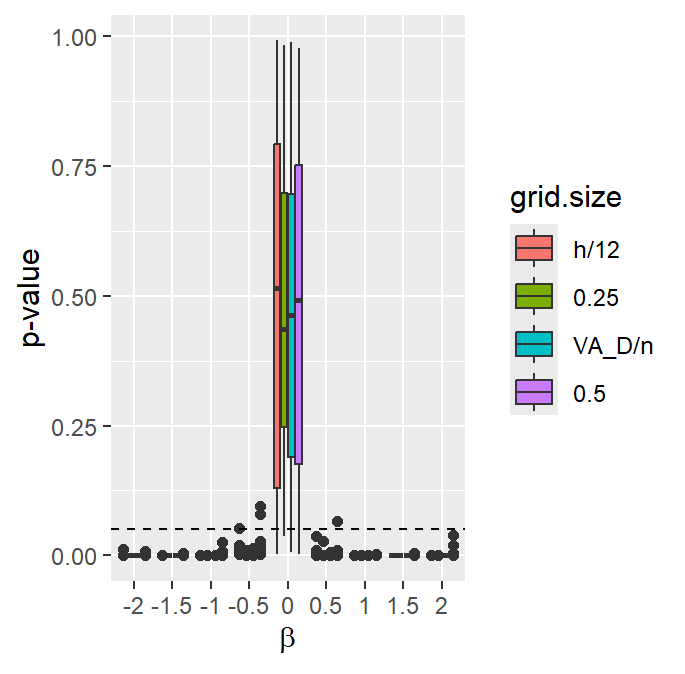}}
\caption{Boxplots of the p-values for the 50 replicas of the MLC test for each value of $\beta$ considered in the data simulation and each value of grid size considered in the test, for the case where the true field mean was $4$, $n=250$. The dashed line represents the 0.05 threshold.  }
	\label{NatarioEtal:figyy3}
\end{figure}

\begin{figure}
	\centerline{
\includegraphics[width=10cm]{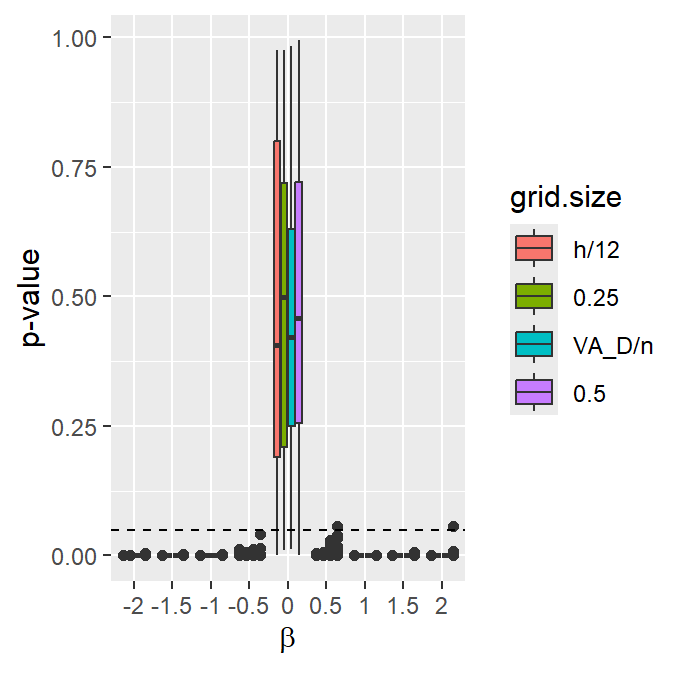}}
\caption{Boxplots of the p-values for the 50 replicas of the MLC test for each value of $\beta$ considered in the data simulation and each value of grid size considered in the test, for the case where the true field mean was distance to coast $dist(x)$, $n=250$. The dashed line represents the 0.05 threshold.   }
	\label{NatarioEtal:figyy4}
\end{figure}

\renewcommand{\arraystretch}{1.1}

\begin{center}
\begin{table}[ht]
\centering
\caption{Proportion of replicas that confirmed preferential sampling on the MCL test (p-value $<~0.05$ ), for different values of $\beta$ used to generate data and different values of grid sizes for the test, and for the case where the true field mean was $4$ and $n=250$.}
\begin{tabular}{ccccc}
  \hline
  & \multicolumn{4}{c}{Grid size} \\ \cline{2-5}
$\beta$ & 0.5 & 0.25 & $h \over 12$ & $\sqrt{\mbox{Area}_{\mathcal{D}}\over n}$ \\
\hline
-2 & 1.00 & 1.00 & 1.00 & 1.00 \\
  -1.5 & 1.00 & 1.00 & 1.00 & 1.00 \\
  -1 & 1.00 & 1.00 & 1.00 & 1.00 \\
  -0.5 & 0.96 & 1.00 & 0.98 & 1.00 \\
  0 & 0.08 & 0.02 & 0.14 & 0.08 \\
  0.5 & 0.98 & 1.00 & 1.00 & 1.00 \\
  1 & 1.00 & 1.00 & 1.00 & 1.00 \\
  1.5 & 1.00 & 1.00 & 1.00 & 1.00 \\
  2 & 1.00 & 1.00 & 1.00 & 1.00 \\   \hline
\end{tabular}
\label{natariocarvalho:tablexx3}
\end{table}
\end{center}

\renewcommand{\arraystretch}{1.1}

\vspace{-1cm}

\begin{center}
\begin{table}[ht]
\centering
\caption{Proportion of replicas that confirmed  preferential sampling on the MCL test (p-value $<~0.05$ ), for different values of $\beta$ used to generate data and different values of grid sizes for the test, and for the case where the true field mean  was distance to coast $dist(x)$ and $n=250$.}
\begin{tabular}{ccccc}
  \hline
  & \multicolumn{4}{c}{Grid size} \\ \cline{2-5}
$\beta$ & 0.5 & 0.25 & $h \over 12$ & $\sqrt{\mbox{Area}_{\mathcal{D}}\over n}$ \\
\hline
-2 & 1.00 & 1.00 & 1.00 & 1.00 \\
  -1.5 & 1.00 & 1.00 & 1.00 & 1.00 \\
  -1 & 1.00 & 1.00 & 1.00 & 1.00 \\
  -0.5 & 1.00 & 1.00 & 1.00 & 1.00 \\
  0 & 0.06 & 0.04 & 0.12 & 0.06 \\
  0.5 & 0.98 & 1.00 & 1.00 & 1.00 \\
  1 & 1.00 & 1.00 & 1.00 & 1.00 \\
  1.5 & 1.00 & 1.00 & 1.00 & 1.00 \\
  2 & 0.98 & 1.00 & 1.00 & 1.00 \\      \hline
\end{tabular}
\label{natariocarvalho:tablexx4}
\end{table}
\end{center}

\begin{figure}
	\centerline{
\includegraphics[width=10cm]{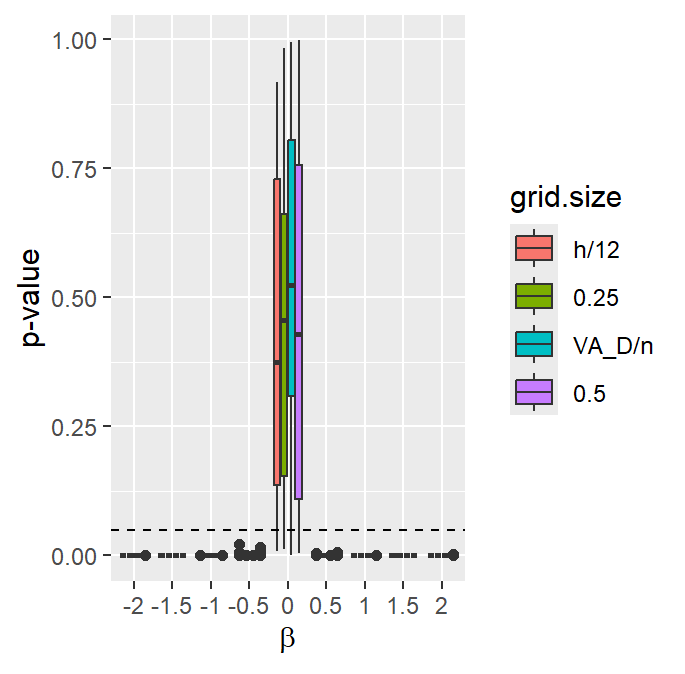}}
\caption{Boxplots of the p-values for the 50 replicas of the MLC test for each value of $\beta$ considered in the data simulation and each value of grid size considered in the test, for the case where the true field mean was $4$, $n=500$. The dashed line represents the 0.05 threshold.  }
	\label{NatarioEtal:figyy5}
\end{figure}

\begin{figure}
	\centerline{
\includegraphics[width=10cm]{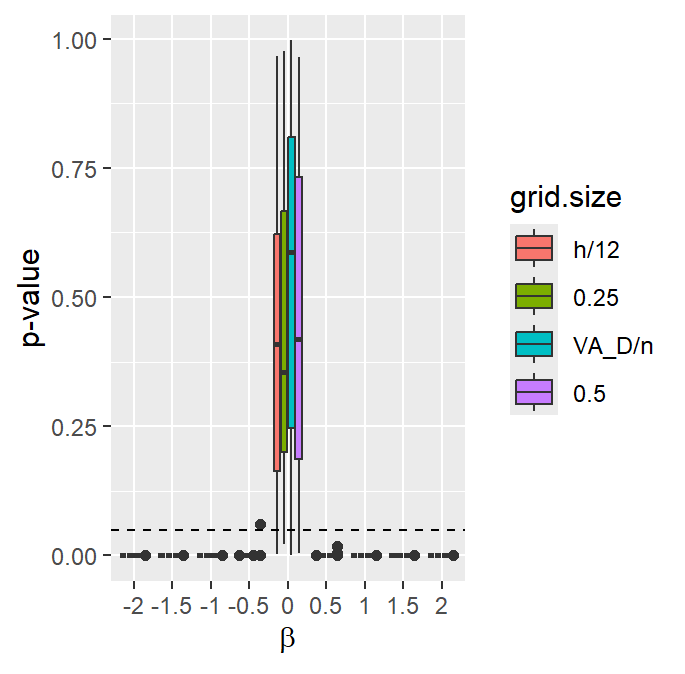}}
\caption{Boxplots of the p-values for the 50 replicas of the MLC test for each value of $\beta$ considered in the data simulation and each value of grid size considered in the test, for the case where the true field mean was distance to coast $dist(x)$, $n=500$. The dashed line represents the 0.05 threshold.   }
	\label{NatarioEtal:figyy6}
\end{figure}

\renewcommand{\arraystretch}{1.1}

\begin{center}
\begin{table}[ht]
\centering
\caption{Proportion of replicas that confirmed preferential sampling on the MCL test (p-value $<~0.05$ ), for different values of $\beta$ used to generate data and different values of grid sizes for the test, and for the case where the true field mean was $4$ and $n=500$.}
\begin{tabular}{ccccc}
  \hline
  & \multicolumn{4}{c}{Grid size} \\ \cline{2-5}
$\beta$ & 0.5 & 0.25 & $h \over 12$ & $\sqrt{\mbox{Area}_{\mathcal{D}}\over n}$ \\
\hline
-2 & 1.00 & 1.00 & 1.00 & 1.00 \\
  -1.5 & 1.00 & 1.00 & 1.00 & 1.00 \\
  -1 & 1.00 & 1.00 & 1.00 & 1.00 \\
  -0.5 & 1.00 & 1.00 & 1.00 & 1.00 \\
  0 & 0.14 & 0.10 & 0.16 & 0.06 \\
  0.5 & 1.00 & 1.00 & 1.00 & 1.00 \\
  1 & 1.00 & 1.00 & 1.00 & 1.00 \\
  1.5 & 1.00 & 1.00 & 1.00 & 1.00 \\
  2 & 1.00 & 1.00 & 1.00 & 1.00 \\    \hline
\end{tabular}
\label{natariocarvalho:tablexx5}
\end{table}
\end{center}

\renewcommand{\arraystretch}{1.1}

\vspace{-1cm}

\begin{center}
\begin{table}[ht]
\centering
\caption{Proportion of replicas that confirmed  preferential sampling on the MCL test (p-value $<~0.05$ ), for different values of $\beta$ used to generate data and different values of grid sizes for the test, and for the case where the true field mean  was distance to coast $dist(x)$ and $n=500$.}
\begin{tabular}{ccccc}
  \hline
  & \multicolumn{4}{c}{Grid size} \\ \cline{2-5}
$\beta$ & 0.5 & 0.25 & $h \over 12$ & $\sqrt{\mbox{Area}_{\mathcal{D}}\over n}$ \\
\hline
-2 & 1.00 & 1.00 & 1.00 & 1.00 \\
  -1.5 & 1.00 & 1.00 & 1.00 & 1.00 \\
  -1 & 1.00 & 1.00 & 1.00 & 1.00 \\
  -0.5 & 0.98 & 1.00 & 1.00 & 1.00 \\
  0 & 0.12 & 0.08 & 0.16 & 0.06 \\
  0.5 & 1.00 & 1.00 & 1.00 & 1.00 \\
  1 & 1.00 & 1.00 & 1.00 & 1.00 \\
  1.5 & 1.00 & 1.00 & 1.00 & 1.00 \\
  2 & 1.00 & 1.00 & 1.00 & 1.00 \\      \hline
\end{tabular}
\label{natariocarvalho:tablexx6}
\end{table}
\end{center}

\clearpage

\bibliographystyle{unsrtnat}
\bibliography{sn-bibliographyINEtal}

@article{BestRoberts1975,
 author = {Best, D. J.  and  Roberts, D. E.},
 journal = {Journal of the Royal Statistical Society. Series C (Applied Statistics)},
 number = {3},
 pages = {377-379},
 title = {Algorithm AS 89: The Upper Tail Probabilities of Spearman's Rho},
 volume = {24},
 year = {1975},
 doi={https://doi.org/10.2307/2347111}
}

@book{BlangiardoCameletti2015,
address = {Chichester, West Sussex},
author = {Blangiardo, Marta and Cameletti, Michela},
isbn = {9781118326558},
language = {English},
month = {jan},
publisher = {Wiley},
title = {{Spatial and Spatio-temporal Bayesian Models with R - INLA}},
year = {2015},
doi = {10.1002/9781118950203}
}

@article{CliffordEtal1989,
 author = {Clifford, P.  and  Richardson, S. and Hemon, D.},
 journal = {Biometrics},
 number = {1},
 pages = {123-134},
 title = {Assessing the Significance of the Correlation between Two Spatial Processes},
 volume = {45},
 year = {1989},
 doi={https://doi.org/10.2307/2532039}
}

@book{DiggleRibeiro07,
	author = {Diggle, Peter and Ribeiro, Paulo Justiniano},
	title = {Model-Based Geostatistics},
    address	= "New York",
	year = {2007},
    series = {Springer Series in Statistics},
    publisher = {Springer}
}

@article{DiggleEtal2010,
title = {Geostatistical Inference Under Preferential Sampling},
journal = {Journal of the Royal Statistical Society Series C: Applied Statistics},
volume = {59},
pages = {191-232},
year = {2010},
doi = {https://doi.org/10.1111/j.1467-9876.2009.00701.x},
author = {Diggle, Peter J. and Menezes, Raquel and Su, Ting-li},
}

@article{Dinsdale2019,
title = {Methods for Preferential Sampling in Geostatistics},
journal = {Journal of the Royal Statistical Society Series C: Applied Statistics},
volume = {68},
pages = {181–198},
year = {2019},
doi = {https://doi.org/10.1111/rssc.12286},
author = {Dinsdale, Daniel and Salibian-Barrera, Matias },
}

@article{DuncanEtal2014,
title = {A Spatially Explicit Approach to the Study of Socio-Demographic Inequality in the Spatial Distribution of Trees across Boston Neighborhoods},
journal = {Spatial Demography},
volume = {2},
pages = {1–29},
year = {2014},
doi = {https://doi.org/10.1007/BF03354902},
author = {Duncan, D.T. and Kawachi, I. and Kum, S. and {Et al.}}
}

@article{GelfandEtal12,
title = {On the effect of preferential sampling in spatial prediction},
journal = {Environmetrics},
volume = {23},
pages = {565-578},
year = {2012},
doi = {https://doi.org/10.1002/env.2169},
author = {Gelfand, Alan E. and Sahu, Sujit K. and Holland, David M.},
}

@article{Guan2006,
title={Tests for Independence between Marks and Points of a Marked Point Process},
author={Guan, Yongtao},
journal={Biometrics},
volume={62},
number={1},
pages={126-134},
year={2006},
publisher={Wiley, International Biometric Society},
doi = {https://doi.org/10.1111/j.1541-0420.2005.00395.x}
}

@article{GuanAfshartous2007,
title = {Test for independence between marks and points of marked point processes: a subsampling approach},
journal = {Environmental and Ecological Statistics},
volume = {14},
pages = {101-111},
year = {2007},
doi = {https://doi.org/10.1007/s10651-007-0010-7},
author = {Guan, Yongtao and Afshartous, David R}
}

@article{Illian2012toolbox,
title={A toolbox for fitting complex spatial point process models using integrated nested Laplace approximation (INLA)},
author={Illian, Janine B and S{\o}rbye, Sigrunn H and Rue, H{\aa}vard},
journal={The Annals of Applied Statistics},
pages={1499--1530},
year={2012},
publisher={JSTOR},
doi = {https://doi.org/10.1214/11-AOAS530}
}

@article{Illian2013,
title={Fitting complex ecological point process models with integrated nested Laplace approximation},
author={Illian, Janine B and Martino, Sara and S{\o}rbye, Sigrunn H and Gallego-Fern{\'a}ndez, Juan B and
Zunzunegui, Mar{\'\i}a and Esquivias, M Paz and Travis, Justin MJ},
journal={Methods in Ecology and Evolution},
volume={4},
number={4},
pages={305--315},
year={2013},
publisher={Wiley Online Library},
doi = {https://doi.org/10.1111/2041-210x.12017}
}

@article{Kass1995,
title={Bayes Factors},
author={Kass, R.E. and Raftery, A.E.},
journal={Journal of the American Statistical Association},
volume={90},
number={430},
pages={773--795},
year={1995},
publisher={Wiley Online Library},
doi = {https://doi.org/10.1080/01621459.1995.10476572}
}

@book{Kendall1948Vol1,
  author    = {Kendall, Maurice G.},
  title     = {The Advanced Theory of Statistics, Vol. 1},
  publisher = {Charles Griffin},
  address   = {London},
  year      = {1948}
}

@book{Kendall1961Vol2,
  author    = {Kendall, Maurice G.  and Stuart, Alan },
  title     = {The Advanced Theory of Statistics, Volume 2: Inference and Relationship},
  publisher = {Charles Griffin},
  year      = {1961}
}

@Inbook{Natario2024,
author="Nat{\'a}rio, Isabel",
editor="Chen, Ding-Geng
and Coelho, Carlos A.",
title="Hypothesis Testing Within Bayesian Inference",
bookTitle="Biostatistics Modeling and Public Health Applications: Study Design and Analysis Methodology in Health Sciences, Volume 1",
year="2024",
publisher="Springer Nature Switzerland",
address="Cham",
pages="29-43",
doi= {https://doi.org/10.1007/978-3-031-69690-9_2}
}

@article{Pennino2019,
  title={Accounting for preferential sampling in species distribution models},
  author={Pennino, Maria Grazia and Paradinas, Iosu and Illian, Janine B and Mu{\~n}oz, Facundo and Bellido,
  Jos{\'e} Mar{\'\i}a and L{\'o}pez-Qu{\'\i}lez, Antonio and Conesa, David},
  journal={Ecology and evolution},
  volume={9},
  number={1},
  pages={653--663},
  year={2019},
  publisher={Wiley Online Library},
  doi = {https://doi.org/10.1002/ece3.4789}
}

@article{Raeisi2021,
title = {A spatio-temporal multi-scale model for Geyer saturation point process: Application to forest fire occurrences},
author = {Raeisi, Morteza  and Bonneu, Florent and Gabriel, Edith },
journal = {Spatial Statistics},
volume = {41},
pages = {100492},
year = {2021},
doi = {https://doi.org/10.1016/j.spasta.2021.100492}
}

@article{ShaddickZidek2014,
title = {A case study in preferential sampling: Long term monitoring of air pollution in the UK},
journal = {Spatial Statistics},
volume = {9},
pages = {51-65},
year = {2014},
doi = {https://doi.org/10.1016/j.spasta.2014.03.008},
author = {Shaddick,Gavin  and Zidek, James V.},
}

@article{Schlather2004,
    author = {Schlather, Martin and Ribeiro, Paulo J., Jr and Diggle, Peter J.},
    title = {Detecting Dependence Between Marks and Locations of Marked Point Processes},
    journal = {Journal of the Royal Statistical Society Series B: Statistical Methodology},
    volume = {66},
    number = {1},
    pages = {79-93},
    year = {2003},
    doi = {10.1046/j.1369-7412.2003.05343.x}
}

@InProceedings{SimoesEtal2023,
author={Sim{\~o}es, Paula and Carvalho, M. Luc{\'i}lia and Figueiredo, Ivone and Monteiro, Andreia and Nat{\'a}rio, Isabel},
editor={Gervasi, Osvaldo and Murgante, Beniamino and Rocha, Ana Maria A. C. and Garau, Chiara and Scorza, Francesco and Karaca, Yeliz
and Torre, Carmelo M.},
title={Black Scabbardfish Species Distribution: Geostatistical Inference Under Preferential Sampling},
booktitle={Computational Science and Its Applications -- ICCSA 2023 Workshops},
year={2023},
publisher={Springer Nature Switzerland},
address={Cham},
pages={303-314},
doi={https://doi.org/10.1007/978-3-031-37108-0_19}
}

@article{Spearman1904,
 author = {Spearman, C.},
 title = {The Proof and Measurement of Association between Two Things},
 journal = {The American Journal of Psychology},
 volume = {15},
 pages = {72-101},
 year = {1904},
 publisher = {University of Illinois Press},
 doi={https://doi.org/10.2307/1412159}
}

@article{vanDoorn2020,
author = {van Doorn, J.  and Ly, A.  and Marsman, M.  and Wagenmakers, E.-J.},
title = {Bayesian rank-based hypothesis testing for the rank sum test, the signed rank test, and Spearman's $\rho$},
journal = {Journal of Applied Statistics},
volume = {47},
number = {16},
pages = {2984-3006},
year = {2020},
publisher = {Taylor \& Francis},
doi = {https://doi.org/10.1080/02664763.2019.1709053}
}

@article{Watson2021,
title = {A perceptron for detecting the preferential sampling of locations and times chosen to monitor a spatio-temporal process},
journal = {Spatial Statistics},
volume = {43},
pages = {100500},
year = {2021},
doi = {https://doi.org/10.1016/j.spasta.2021.100500},
author = {Watson, Joe}
}






\end{document}